\documentclass[%
 reprint,
 amsmath,amssymb,
 aps,
]{revtex4-2}

\usepackage{graphicx}
\usepackage{dcolumn}
\usepackage{bm}
\usepackage{booktabs}       
\usepackage{siunitx}        
\usepackage{adjustbox}      

\begin{document}

\preprint{APS/123-QED}

\title{Exploring Satellite Quantum Key Distribution under Atmospheric Constraints}
\author{Aditya Ajith}
 \author{S. Saravana Veni}%
 \email{s\textunderscore saravanaveni@cb.amrita.edu} 
\affiliation{%
 Department of Physics, Amrita School of Physical Sciences, Amrita Vishwa Vidyapeetham, Coimbatore - 641112, Tamil Nadu, India. 
}%

\date{\today}

\begin{abstract}
Satellite Quantum Key Distribution creates a pathway for secure global communication with a level of security that is peerless. However, ground-to-satellite Quantum Key Distribution links are degraded due to the atmospheric turbulence. This paper gives a numerical framework using angular spectrum propagation, Hufnagel-Valley model of turbulence and Von Karman phase screens and takes into account the static losses introduced due to the absorption of the beam by the different elements and compounds in the atmosphere like $O_2$, $CO_2$ and $H_2O$. This simulation propagates a Gaussian beam step by step through the atmosphere, where we incorporate the phase distortions using phase screens based on standard $C_n^2$ profiles which take into account losses such as scintillation and beam wander. We simulate the BB84 protocol with decoy states for added security. The results of the simulation quantify the expected link budget and Secure key rates over a range of distances to measure the viability of Free Space Optical Quantum Key Distribution links at different distances.
\end{abstract}

\maketitle


\section{\label{sec:level1}Introduction}
Quantum key distribution stands at the forefront of quantum computation and has emerged as one of the first quantum technologies to have evolved from theoretical research to practical implementation. Originally demonstrated in free space propagation under laboratory conditions \cite{bennett1992experimental} 
after which it moved on larger scale links using optical fibres. In the realm of free space propagation, secure key exchanges over free space link have taken place over 144 km \cite{ursin2007entanglement}. 
This field has its foundations in the year of 1984, when Giles Brassard and Charles Bennet put forward the first QKD protocol which is now known as the BB84 protocol \cite{bennett2014quantum}. Since then this field has undergone substantial development. Protocols leveraging phenomena like entanglement, have only increased the practicality and security of QKD \cite{ekert1991quantum}. One of the challenges in this field is that of the distribution of the quantum bits or qubits carrying the bit information to the users. One of the solutions is to use the satellites to either directly distribute the keys or act as trusted nodes to relay the qubits the users, thus facilitating key exchange \cite{bedington2017progress,liao2017satellite}. This method of key distribution relies on Free Space Optical (FSO) communication which uses laser beams to send the data through the atmosphere. However, FSO communication comes with its own set of drawbacks, that is the beams are prone to degradation which are due to reasons such as: Turbulence, beam spreading, beam wander, scattering and absorption \cite{majumdar2010free} \cite{andrews2019field}. This can lead to decoherence, a quantum phenomena where the quantum state loses its quantum properties due to interaction with the environment \cite{pirandola2020advances}, which means it can lose the value of the bit its carrying or change its value giving a wrong reading. This can affect the fidelity or the efficiency of the system. While atmospheric effects can degrade the signal, the main avenue we have to focus on especially for BB84 is the loss and error since photon loss is a bigger issue than decoherence \cite{bedington2017progress}. Which is why it is important to accurately model the atmosphere and simulate the loss offered by the channel before deploying this to the real world and testing. This melding of quantum mechanics to information theory and cryptography has reshaped our understanding of data and information; from something abstract to physical in nature. By exploiting the non-classical nature and properties of photons such as polarization, entanglement, and interference, QKD gives us a level of security that unmatched by classical methods. Theorems such as No-cloning theorem and measurement of quantum states collapsing the system ensures that if there is an eavesdropper the users will be informed of it \cite{mafu2018security,nurhadi2018quantum,wootters1982single,dieks1982communication}. This paper provides a simulation framework for a ground to space QKD link using a wave optics approach. We simulate turbulence induced phase perturbations, static beam attenuation and beam propagation. After which we calculate the Quantum Bit Error Rate (QBER) and Secure Key Rate (SKR), which are validated against theoretical limits.
\subsection{Beam Propagation Theory}
A high fidelity wave opics is executed to trace the atmospheric journey of a laser beam, capturing its transformation to a satellite mounted receiver in orbit. To evaluate the performance of a ground to satellite QKD link, this simulation initializes a Gaussian beam at the initial point, i.e., the ground station, which is defined with a wavelength $\lambda$ (810 nm) and an appropriate beam waist $W_0$ that is propagated through the atmosphere \cite{sidhu2021advances}.
To propagate the beam over a distance $L$, employed here is an Angular Spectrum Method (ASM) that uses Fast Fourier Transforms (FFT) to take the complex field $E(x,y,z)$ into its angular spectrum $A(k_x,k_y,z)$ in the spatial frequency domain $(k_x,k_y)$, where $k_x, k_y$ are the spatial frequencies. The field is then multiplied with the ASM Propagator $H(k_x,k_y,dz)$ \cite{goodman2005introduction}
\begin{equation}
 H(k_x, k_y, dz) = \exp\left(i k_z dz\right) = \exp\left(i \sqrt{k^2 - k_x^2 - k_y^2} dz\right).
\end{equation}
Here $k=2\pi/\lambda$ is the wavenumber, and $k_z$ represents the propagation constant along the z-axis. An inverse FFT brings back the beam to the spatial domain.
The complex field is represented as a $N \times N$ grid with a spacing of $\delta$ (e.g., $1 cm$) and is propagated through the entire distance with a step size of $dz$. After the inverse FFT, the beam is propagated a distance of $z+dz$. The constants are chosen such that the ASM sampling factor is sufficiently large. The ASM Sampling factor is given by
\begin{equation}
S=\frac{N\delta^2}{\lambda dz}.
\end{equation}
This is done so that we can minimize the numerical aliasing artifacts for the angular spread encountered during beam propagation.
\subsection{Turbulence Theory}
Atmospheric turbulence introduces random phase aberrations and intensity fluctuations.
Employed here is the Hufnagel-Valley (HV) Model of turbulence to calculate the strength of turbulence and gives the altitude-dependent refractive index structure factor which is characterized by $C_n^2(h)$ \cite{andrews2019field} \cite{zhou2023atmospheric}:
\begin{eqnarray}
 C_n^2(h) &=& 0.00594 \left( \frac{v}{27} \right)^2 (10^{-5}h)^{10} \exp\left(-\frac{h}{1000}\right) \nonumber \\
 & & + 2.7 \times 10^{-16} \exp\left(-\frac{h}{1500}\right) \nonumber \\
 & & + A_0 \exp\left(-\frac{h}{100}\right).
 \label{eq:cn2}
\end{eqnarray}
Here $v$ stands for root mean square wind velocity, $h$ stands for the altitude, and $A_0$ is the ground level turbulence strength.
The effect of turbulence over a path is given by the Fried parameter $r_0$ \cite{fried1966optical} which is calculated by integrating $C_n^2(h)$ along the path length. A larger $r_0$ means weaker turbulence:
\begin{equation}
r_0 = \left[0.423 k^2 \sec(\theta) \int_{\text{path}} C_n^2(h) dh\right]^{-3/5}.
\label{eq:r0}
\end{equation}
Here $\theta$ is the zenith angle.
\begin{figure*}
    \centering
    \includegraphics[width=1\linewidth]{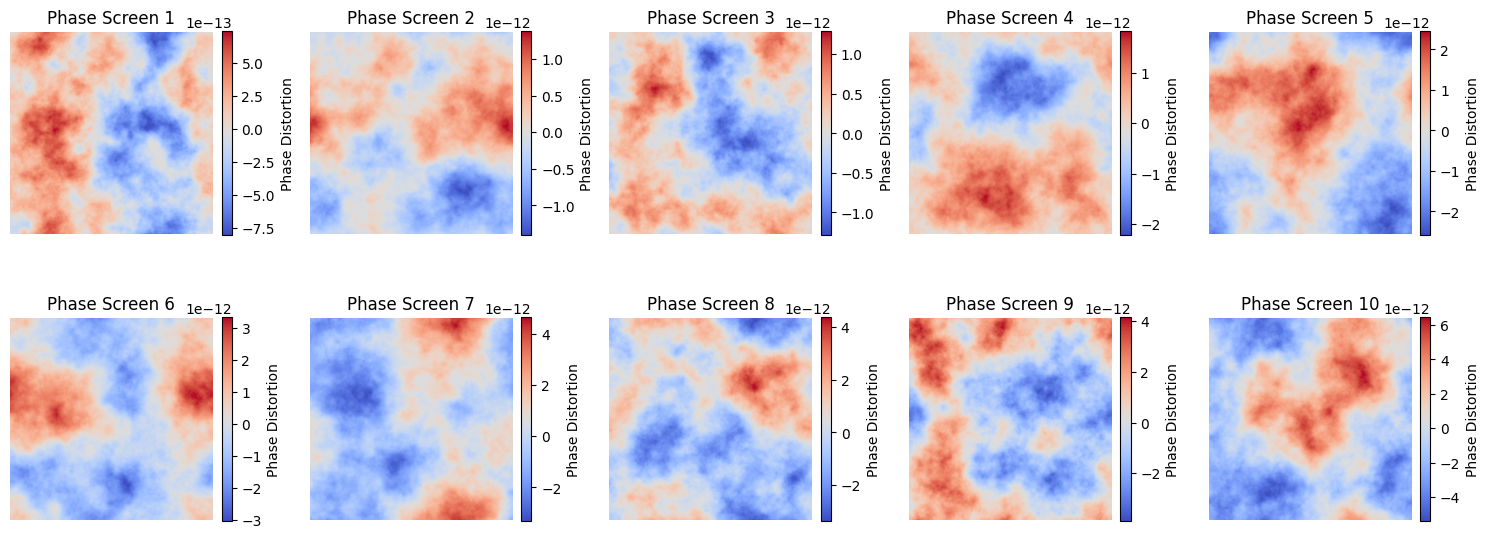}
    \caption{This figure displays random realizations of the phase screens that are generated to model the turbulent effects of the atmosphere over a single propagation step. }
    \label{fig:phasescreen}
\end{figure*}
The turbulence is simulated using random phase screens as seen in Fig. \ref{fig:phasescreen}. The phase screens $\phi(x,y)$ are generated based on the Von Karman power spectrum. The Von Karman power spectral density formula describes how the phase variations are distributed across different spatial scales, incorporating the inner scale of turbulence $l_0$ and the outer scale $L_0$. These values define the size of the eddies \cite{charnotskii2020comparison}. The outer scale $L_0$ describes the largest size of the turbulent eddy and the inner scale $l_0$ gives us the smallest size of the turbulent eddies; this determines how fine-grained the turbulence is. The effect of turbulence is simulated using random phase screens applied multiplicatively to the complex optical field at each step $dz$. As the beam propagates and is subject to further phase screens, it naturally gives rise to scintillation in the beam profile; scintillations are intensity fluctuations on the beam. I also take into account beam wander, which takes place when the beam encounters larger eddies which thereby cause a full change in direction of the beam due to the large-scale phase tilts in the beam wavefront. This is simulated by calculating the expected angle of arrival using $r_0$ and the aperture size, and adding a random lateral shift to the complex grid at each step $dz$ \cite{andrews2019field}.
\subsection{Static Losses}
Beyond turbulence, the atmosphere causes deterministic power loss through absorption by gases (primarily $H_2O, O_2, O_3$ in the relevant bands) and scattering by air molecules (Rayleigh scattering) and aerosols (Mie scattering). These combined effects determine the static atmospheric path transmittance $\eta_{\text{atm}}$ for a specific path geometry and wavelength. This loss can be modelled using the Beer's Law.
\[\eta_{atte}=\frac{I(L)}{I(0)}=e^{-\beta_{ext}(h,\lambda)L}
    \]
where $\beta_{ext}$ is the extinction coefficient \cite{behera2024estimating}.
To evaluate the static atmospheric losses, an initial attempt was made using the Python-based radiative transfer interface Py6S \cite{wilson2013py6s}. However, the outcomes diverged from expected values for the specific location under study. This discrepancy is likely attributable to the default aerosol models and atmospheric profiles employed in Py6S, which differ from those used in MODTRAN, or due to variations in how gaseous absorption is treated for high-altitude upward propagation paths. Consequently, for this parameter, I have relied on highly specialized MODTRAN-derived results reported in the literature \cite{behera2024estimating}. The calculated transmittance quantifies the proportion of light that traverses the atmosphere without being scattered or absorbed by its constituents.
By utilizing HITRAN’s Application Programming Interface (HAPI), I computed the transmittance across a broad spectral range, enabling the identification of specific wavelength bands commonly employed in Free Space Optical (FSO) communication. The HITRAN database provides an extensive repository of spectroscopic parameters for atmospheric gases, which allows for precise modeling of absorption features critical to high-fidelity atmospheric transmittance analysis \cite{kochanov2016hitran}.
\begin{figure}
    \centering
    \includegraphics[width=0.8\linewidth]{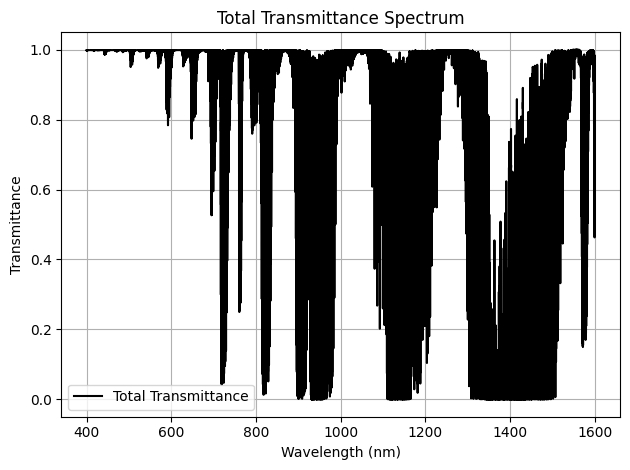}
    \caption{This graph displays the calculated total atmospheric transmittance as a function of wavelength (nm) over the spectral range of 400 nm to 1600 nm.}
    \label{}
\end{figure}
To evaluate the efficiency of the quantum link, employed in the code using BB84 protocol \cite{bennett2014quantum} with decoy states \cite{mizutani2025protocol}. This protocol enables the users (Alice and Bob) establish a secure shared key, with its security backed by the laws of quantum mechanics. This works by encoding the bits onto quantum states, in this case the polarization state of the photons, which are decided by mutually unbiased bases.
An ideal BB84 protocol would use single photons with the bit encoded onto the polarization state. However, due to current technological limits, on-demand single-photon sources are not viable. So, practical QKD links use Weak Coherent Pulses (WCP) \cite{bozzio2022enhancing}. These are produced by attenuating the laser beam. Photons follow a Poisson distribution \cite{loudon2000quantum}, which means that the pulse can have 0, 1 or more photons per pulse. This is characterized by the mean photon number $\mu$. This is a vulnerability as this can open up the link to Photon Number Splitting (PNS) attacks; this is when the eavesdropper can take a few photons from each pulse and get the secret key without alerting the users to this breach in security \cite{lutkenhaus2002quantum}.
To battle this vulnerability we use decoy states \cite{hwang2003quantum,lo2005decoy,wang2005beating}. Now when Alice sends a pulse to Bob it can have a random photon number, for example, $\mu_{\text{decoy}} \approx 0.1$ for a decoy state and $\mu_{\text{signal}} \approx 0.5$ for a signal state. To an eavesdropper they have no possibility of knowing whether the pulse they get is a decoy state or a signal state. Then by comparing detection rates and error rates across signal and decoy states, Alice and Bob can estimate the fraction of detections due to single-photon pulses and the upper bounds on the attacker's knowledge.
\subsection{Gain and QBER calculation}
We model a WCP QKD system using BB84 and the expected gain ($Q_{\mu}$) and QBER ($E_{\mu}$) are the key perofrmance indicators. The quantity \( T_{\text{path,avg}} \), representing the average path transmittance, is derived by averaging the instantaneous transmittance values over multiple optical pulses \cite{pirandola2020advances}. This metric reflects the proportion of optical power that effectively propagates through a turbulent atmospheric channel. In addition to this, several other crucial system parameters are incorporated to determine the Quantum Bit Error Rate (QBER).
$\mu$ is Mean photon number per pulse at the source, $\eta_{t, \text{optical}} and \eta_{r, \text{optical}}$ are  Optical efficiencies of the transmitter and receiver respectively. $\eta_d$ is Detector efficiency (quantum efficiency),
$Y_0$ is Dark count probability per detector per coincidence window, $e_{\text{det}}$ is intrinsic detector error probability (probability of an incorrect outcome due to detector noise, not a dark count) and $e_{\text{pol}}$ is Polarization misalignment error. After the total channel efficiency is calculated using this formula,
\begin{equation}
\eta_{\text{total}} = \eta_{t, \text{optical}} \cdot T_{\text{path, avg}} \cdot \eta_{r, \text{optical}},
\end{equation}
we can calculate the average number of photons arriving at the detector for a pulse with mean photon number $\mu$ to be
\begin{equation}
\mu_{\text{det}}=\mu \cdot \eta_{\text{total}} \cdot \eta_d.
\end{equation}
The gain $Q_\mu$ (probability of at least one click at the receiver for a pulse sent with mean photon number $\mu$) is given by:
\begin{equation}
Q_\mu = Y_0 + (1-Y_0)(1 - e^{-\mu_{\text{det}}}).
\label{eq:gain}
\end{equation}
This formula correctly accounts for clicks from signal photons and dark counts.
The QBER $E_\mu$ for pulses sent with mean photon number $\mu$ is the ratio of erroneous clicks to total clicks:
\begin{equation}
E_\mu Q_\mu = e_{\text{err}} (1-Y_0)(1 - e^{-\mu_{\text{det}}}) + \frac{1}{2} Y_0,
\label{eq:qber_num}
\end{equation}
where $e_{\text{err}} = e_{\text{det}} + e_{\text{pol}}$ (assuming $e_{\text{det}}$ is the probability of a photon causing a click in the wrong detector, and $e_{\text{pol}}$ is the polarization misalignment contribution). The $1/2$ factor for $Y_0$ assumes dark counts are equally likely in any detector basis.
Hence
\begin{equation}
E_\mu = \frac{ e_{\text{err}} (1-Y_0)(1 - e^{-\mu_{\text{det}}}) + 0.5 Y_0 }{ Q_\mu }.
\label{eq:qber}
\end{equation}
\subsection{Secure Key Rate calculation}
The ultimate performance metric is the estimated Secure Key Rate (SKR). We use the asymptotic key rate formula for decoy-state BB84 (referencing the principles of GLLP analysis \cite{Gottesman2004Security,scarani2009security}):
\begin{equation}
R_{\text{SKR}} \geq R_{\text{rep}} \cdot P_{\text{sift}} \cdot \left[ Q_1^{\text{L}} (1 - H_2(E_1^{\text{U}})) - Q_\mu f_{\text{EC}} H_2(E_\mu) \right],
\label{eq:skr}
\end{equation}
where $R_{\text{rep}}$ is the pulse repetition rate, $P_{\text{sift}}$ is the sifting factor (typically 0.5 for BB84), $Q_1^{\text{L}}$ is the lower bound on the gain of single-photon pulses, $E_1^{\text{U}}$ is the upper bound on the error rate of single-photon pulses (both estimated using decoy state analysis), $Q_\mu$ is the gain of the signal state, $E_\mu$ is its QBER, $f_{\text{EC}}$ is the efficiency of the error correction algorithm (typically $\geq 1$)\cite{brassard1993secret}, and $H_2(x) = -x \log_2(x) - (1-x) \log_2(1-x)$ is the binary entropy function. The terms $Q_1$ and $E_1$ in the table are estimates for these values.
\section{\label{sec:results}Results and Discussion}
The simulation results presented in Table~\ref{tab:qkd_summary_810nm_part1} and \ref{tab:qkd_summary_810nm_part2} show a clear trend of decreasing secure key rates with increasing satellite distance. This is primarily due to the increase in total signal loss (Loss Total Sig(dB)), which is dominated by the path simulation loss (Loss Path Sim(dB)) and geometric ideal loss (Loss Geom Ideal(dB)) at longer distances. The results were verified using the results from the literature \cite{behera2024estimating}
\begin{figure}
    \centering
    \includegraphics[width=1\linewidth]{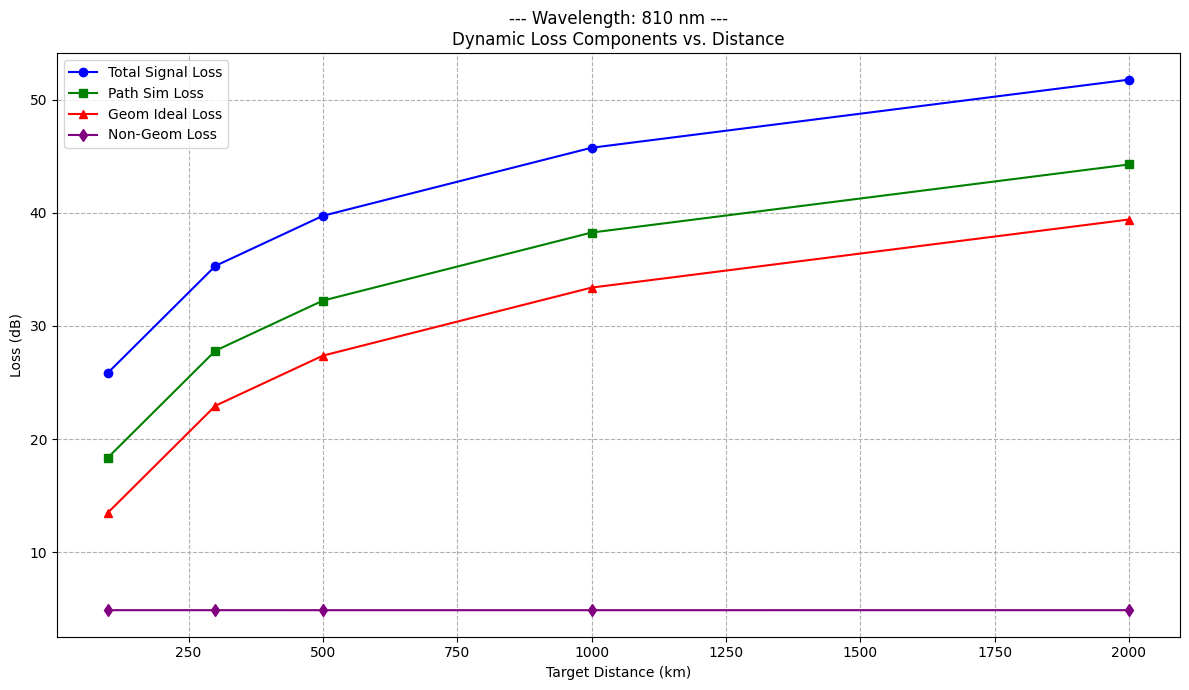}
    \caption{Simulated optical loss components as a function of target distance for an 810 nm ground-to-space uplink. Plotted are the Total Signal Loss (incorporating all system and atmospheric effects), Path Simulation Loss (geometric and turbulence-induced atmospheric loss), Ideal Geometric Loss (vacuum diffraction only), and Non-Geometric Loss (additional loss due to atmospheric effects beyond ideal diffraction).}
    \label{fig:enter-label}
\end{figure}
\begin{table*}[htbp] 
\centering
\caption{Summary of simulated loss metrics and raw gain versus distance for an 810 nm uplink (Part 1).}
\label{tab:qkd_summary_810nm_part1}
\begin{tabular}{
  S[table-format=4.1]    
  S[table-format=2.3]    
  S[table-format=2.3]    
  S[table-format=2.3]    
  S[table-format=1.4e2]  
}
\toprule
\multicolumn{1}{c}{\parbox{1.8cm}{\centering Target Dist \\ (km)}} & 
\multicolumn{1}{c}{\parbox{1.8cm}{\centering Loss Total \\ Sig (dB)}} &
\multicolumn{1}{c}{\parbox{1.8cm}{\centering Loss Path \\ Sim (dB)}} &
\multicolumn{1}{c}{\parbox{1.8cm}{\centering Loss Geom \\ Ideal (dB)}} &
\multicolumn{1}{c}{\parbox{2.2cm}{\centering $T_{\text{raw\_sim\_shape}}$ \\ Gain}} \\ 
\midrule
100.0 & 25.864 & 18.366 & 13.497 & 4.4828e-02 \\
300.0 & 35.316 & 27.818 & 22.949 & 5.0859e-03 \\
500.0 & 39.746 & 32.247 & 27.379 & 1.8341e-03 \\
1000.0 & 45.763 & 38.264 & 33.396 & 4.5888e-04 \\
2000.0 & 51.784 & 44.285 & 39.416 & 1.1470e-04 \\
\bottomrule
\end{tabular}
\end{table*}

\begin{table*}[htbp] 
\centering
\caption{Summary of simulated QKD performance metrics versus distance for an 810 nm uplink (Part 2).}
\label{tab:qkd_summary_810nm_part2}
\begin{tabular}{
  S[table-format=4.1]    
  S[table-format=1.3e2]  
  S[table-format=2.4]    
  S[table-format=1.3e2]  
  S[table-format=1.4]    
  S[table-format=1.3]    
  S[table-format=4.3]    
}
\toprule
\multicolumn{1}{c}{\parbox{1.8cm}{\centering Target Dist \\ (km)}} &
\multicolumn{1}{c}{\parbox{1.5cm}{\centering $\mu$ \\ (sim)}} & 
\multicolumn{1}{c}{\parbox{2cm}{\centering QBER $\mu$ sim \\ (\%)}} &
\multicolumn{1}{c}{\parbox{1.5cm}{\centering $Q_1$ \\ calc}} &
\multicolumn{1}{c}{\parbox{1.8cm}{\centering $E_1$ calc \\ (\%)}} &
\multicolumn{1}{c}{\parbox{1.8cm}{\centering Sift Rate \\ (kbps)}} &
\multicolumn{1}{c}{\parbox{2cm}{\centering Secure Rate \\ (bps)}} \\
\midrule
100.0 & 1.296e-03 & 2.5367 & 2.592e-03 & 2.5188 & 6.480 & 9413.210 \\
300.0 & 1.480e-04 & 2.8209 & 2.945e-04 & 2.6655 & 0.740 & 1044.168 \\
500.0 & 5.401e-05 & 3.3794 & 1.065e-04 & 2.9576 & 0.270 & 360.046 \\
1000.0 & 1.426e-05 & 5.8300 & 2.703e-05 & 4.3036 & 0.071 & 72.641 \\
2000.0 & 4.316e-06 & 13.5064 & 7.131e-06 & 9.3360 & 0.022 & 4.661 \\
\bottomrule
\end{tabular}
\end{table*}

The Fried parameter ($r_0$) remains constant in this simulation setup, implying a consistent turbulence profile assumption across different path lengths, which might be a simplification if the ground station characteristics or zenith angle change significantly for different orbits.
\begin{figure}
       \includegraphics[width=1.0\linewidth]{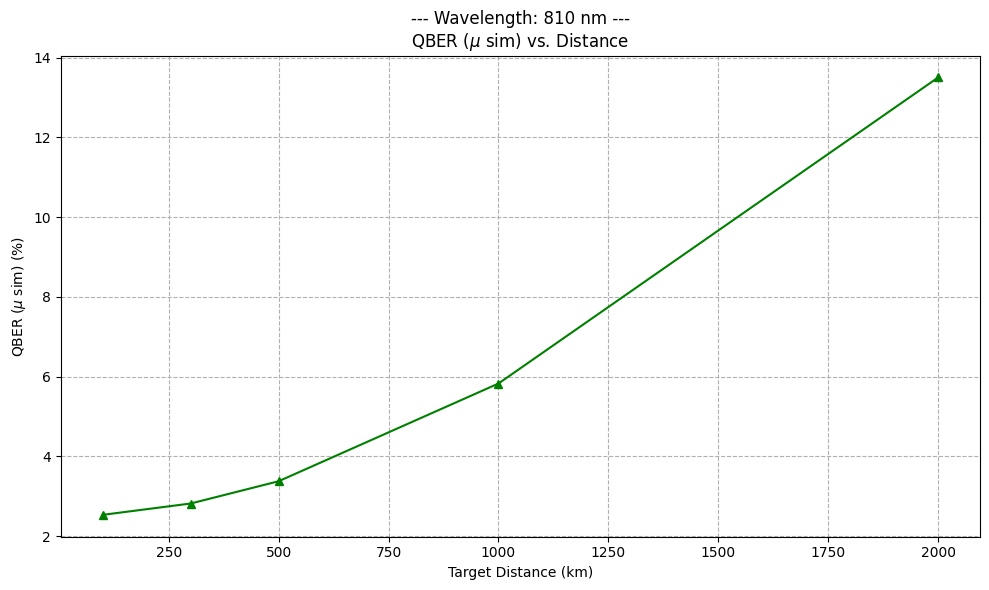}
    \caption{Simulated Quantum Bit Error Rate (QBER, $E_\mu$) for the signal state ($\mu$=0.5) as a function of target distance (km) for the 810 nm ground-to-space QKD link. QBER is calculated based on simulated channel transmittance, dark counts ($Y_0$=$1\times10^{-6}$), and system error probabilities ($e_{detector}$ =0.015.}
    \label{fig:QBER}
\end{figure}
\begin{figure}
       \includegraphics[width=1.0\linewidth]{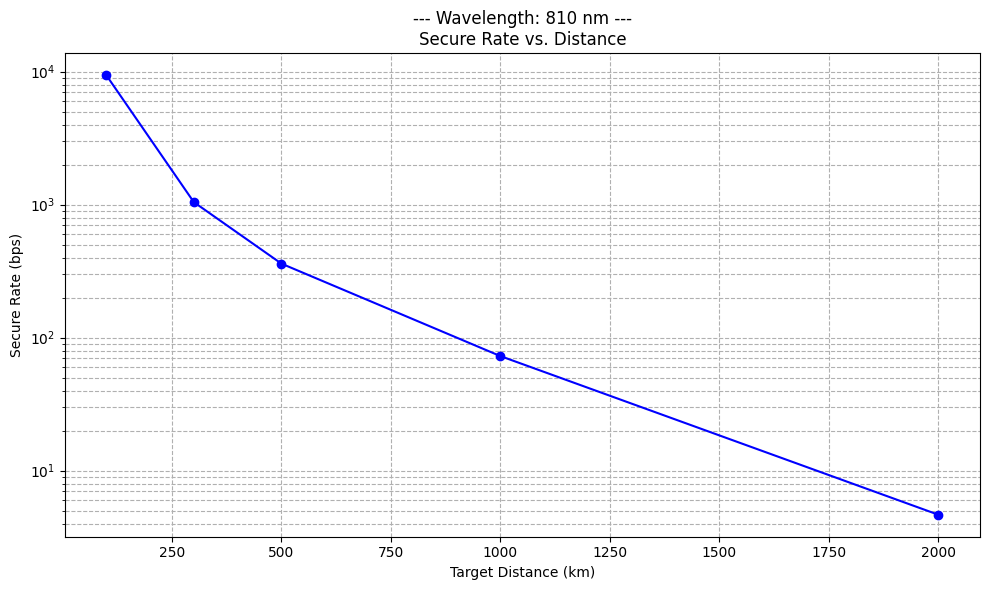}
    \caption{The graph illustrates the calculated asymptotic Secure Key Rate (bps) versus the target propagation distance (km). The SKR is derived using the GLLP security analysis framework for a BB84 protocol with decoy states, based on simulated signal state gain ($Q_\mu$) and QBER ($E_\mu$), estimated single-photon parameters ($Q_1$, $E_1$), a system repetition rate of $R_{\text{rep}}$=$1\times 10^7$ Hz], and an error correction inefficiency factor $f_{\text{EC}}$=1.22. Note the logarithmic scale on the y-axis, highlighting the exponential decrease in SKR with increasing distance due to accumulated channel losses and noise.}
    \label{fig:Secure_rate}
\end{figure}
The QBER ($\mu$ sim) increases with distance , as we can see in fig.\ref{fig:QBER}, exceeding 10\% at 2000 km. This rise in QBER, coupled with sharply falling sift rates, severely limits the achievable secure key rate, dropping from over 9 kbps at 100 km to a mere 4.6 bps at 2000 km for the 810 nm wavelength.
These findings highlight the critical challenge of optical losses in free-space QKD and underscore the necessity for adaptive optics\cite{tyson1996adaptive}, advanced error correction, and potentially higher initial pulse rates or more robust protocols for GEO distances.
\begin{figure*}
 \includegraphics[width=1.0\linewidth]{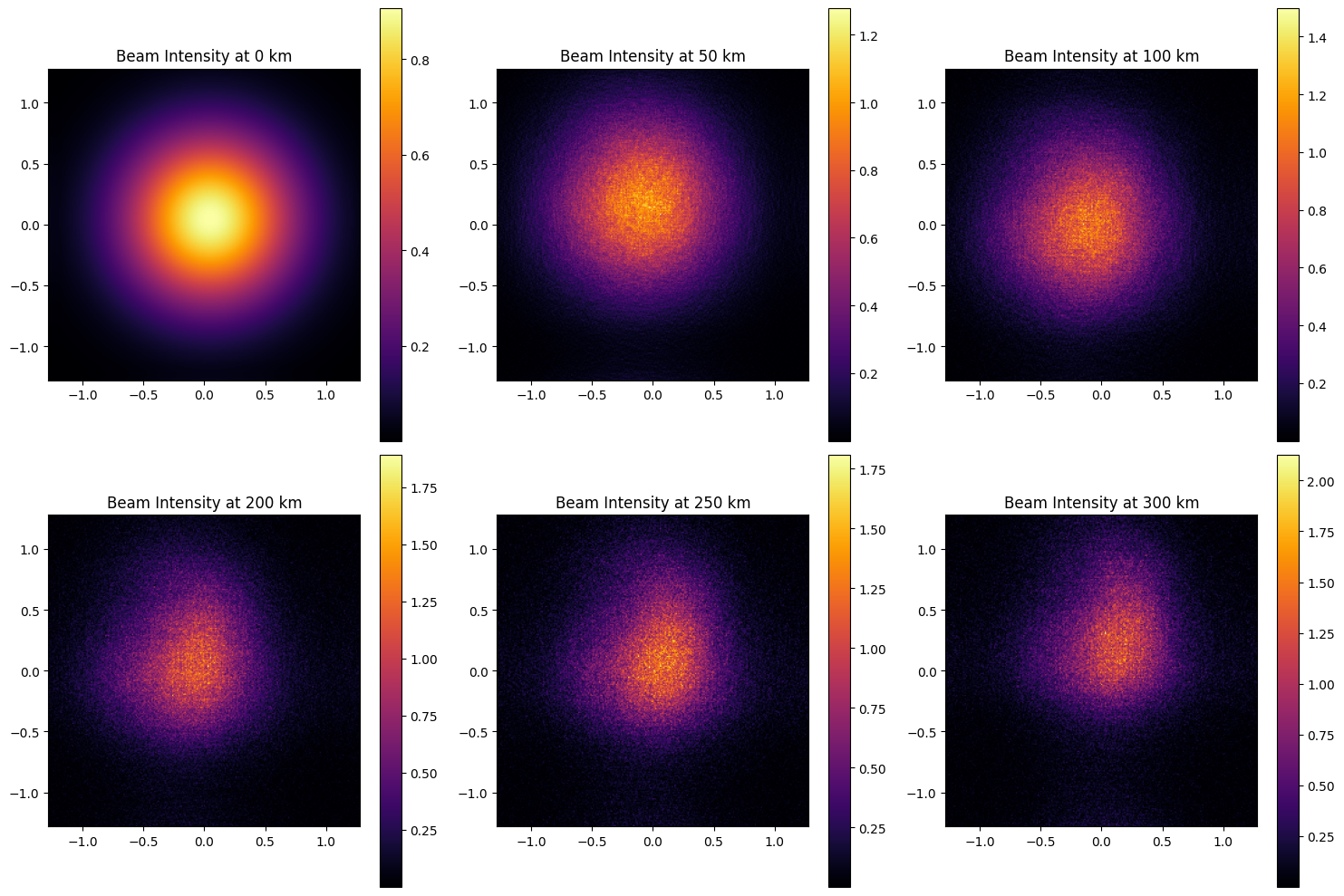}
\caption{An enhanced view of the beam profile capturing the scintillation, beam wander, and attenuation effects of the atmosphere on the beam.}
\label{fig:Beam_profile}
\end{figure*}
\section{\label{sec:conclusion}Conclusion}
This study developed a comprehensive numerical framework to simulate ground-to-satellite Quantum Key Distribution (QKD) links, integrating critical optical effects such as wave optics propagation, atmospheric turbulence modeled via phase screens, and static atmospheric losses. By employing the BB84 protocol augmented with decoy states, the framework enabled a realistic assessment of secure key generation rates across various satellite altitudes, from Low Earth Orbit (LEO) to higher Earth Orbit. 
The simulations highlighted how atmospheric phenomena—specifically scintillation, beam wander, and attenuation—substantially influence the fidelity and efficiency of QKD links. Enhanced beam profile analysis (as shown in Fig. 6) provided visual and quantitative insights into the distortive impacts of the turbulent medium, validating the necessity of incorporating dynamic channel modeling in satellite QKD studies. 
Our results, summarized in Tables I and II, reveal that while LEO-based QKD links yield favorable secure key rates under realistic conditions, the extension to higher orbits introduces pronounced performance degradation due to cumulative losses and increased beam divergence. This performance drop underscores the need for advanced mitigation techniques, such as adaptive optics, higher transmission power, or error-resilient quantum protocols, to ensure viable long-distance quantum communication.

Ultimately, the framework presented here serves as a robust and scalable tool for the evaluation and optimization of satellite QKD architectures. It offers a critical stepping stone toward the realization of global-scale quantum communication networks, bridging theoretical design and practical deployment. Future work will focus on integrating real-time adaptive correction strategies and extending the model to dynamic satellite-ground link scenarios, further enhancing its utility for mission planning and technology development.
\section*{Acknowledgements}
S. Saravana Veni acknowledges Amrita Vishwa Vidyapeetham,
Coimbatore, where this work was supported under Amrita Seed Grant (File
Number: ASG2022141).

\section{Data Availability Statement}
No more existing data were used in this paper.

\end{document}